\documentclass[aps,prd,showpacs,nofootinbib,floats,floatfix,reprint,preprintnumbers,groupedaddress,twocolumn]{revtex4-1}

\usepackage{bm}
\usepackage{latexsym}
\usepackage{dcolumn}
\usepackage{amsmath,amsfonts,amssymb}
\usepackage{graphicx,epsfig}
\usepackage{amsmath}
\usepackage{fancyhdr}
\usepackage{hyperref}
\usepackage{graphicx,epstopdf}
\hypersetup{
	colorlinks   = true, 
	urlcolor     = blue, 
	linkcolor    = blue, 
	citecolor   = red 
}

\def\bea{\begin{eqnarray}}
\def\eea{\end{eqnarray}}
\def\nno{\nonumber}

\begin{document}
\title{Limitations of Pseudo-Newtonian approach in studying the accretion flow around Kerr black hole}
\author{Indu K. Dihingia}\email{i.dihingia@iitg.ac.in}
\author{Santabrata Das}\email{sbdas@iitg.ac.in (Corresponding Author)}
\author{Debaprasad Maity}\email{debu@iitg.ernet.in}
\author{Sayan Chakrabarti}\email{sayan.chakrabarti@iitg.ac.in}

\affiliation{Department of Physics, Indian Institute of Technology Guwahati, Guwahati 781039, Assam, India
}

\date{\today}

\begin{abstract}
    We study the relativistic accretion flow in a generic stationary axisymmetric space-time and obtain an effective potential ($\Phi^{\rm eff}$) that accurately mimics the general relativistic features of Kerr black hole having spin $0\le a_{\rm k} <1$.  Considering the accretion disc to be confined around the equatorial plane of a rotating black hole and using the relativistic equation of state, we examine the properties of the relativistic accretion flow  and compare it with the same obtained form semi-relativistic as well as non-relativistic accretion flows. Towards this, we first investigate the transonic properties of the accretion flow around the rotating black hole where good agreement is observed for relativistic and semi-relativistic flows. Further, we study the non-linearities such as shock waves in accretion flow. Here also we find that the shock properties are in agreement for both relativistic and semi-relativistic flows irrespective of the black hole spin ($a_{\rm k}$), although it deviates significantly for non-relativistic flow. In fact, when the particular shocked solutions are compared for flows with identical outer boundary conditions, the positions of shock transition in relativistic and semi-relativistic flows agree well with deviation of $6-12\%$ for $0 \le a_{\rm k} \le 0.99$, but vast disagreement is observed for non-relativistic flow. In addition, we compare the parameter space (in energy (${\cal E}$) and angular momentum ($\lambda$) plane) for shock to establish the fact that relativistic as well as semi-relativistic accretion flow dynamics do show close agreement irrespective of $a_{\rm k}$ values, whereas non-relativistic flow fails to do so. With these findings, we point out that semi-relativistic flow including $\Phi^{\rm eff}$ satisfactorily mimics the relativistic accretion flows around Kerr black hole. Finally, we discuss the possible implications of this work in the context of dissipative advective accretion flow around Kerr black holes.  

\end{abstract}

\maketitle

\section{Introduction}

The accretion of matter on to black hole is considered to be an essential physical phenomenon as it is regarded to be the principal source of power in microquasars, active galactic nuclei and quasars \citep{Frank-etal2002}.
The overall  description through which one studies the accretion process is the hydrodynamic flow of matter in the background of black hole space-time. Indeed, when inflowing matter approaches towards the horizon, the general relativistic effects become important and due to non-linearity, it is in general difficult to solve the problem. To avoid complexity, therefore, most of the studies of accretion flow around black holes were confined in the Newtonian regime where gravitational effect is taken into account using effective potentials. In practice, while studying accretion dynamics, people conventionally adopt some trial effective potentials known as Pseudo-Newtonian potentials that approximately mimic the general relativistic effects around the black hole. This evidently yields erroneous result particularly when one studies the physical processes in the vicinity of the black hole.
Therefore, the search for an effective potential that accurately describes the space-time geometry around the black hole is very much appealing although it is an age old endeavor in the context of black hole accretion process and in this work, we attempt to do so.

In case of accretion around Schwarzschild black holes, the pseudo-Newtonian potential was first proposed by \citet{Paczynsky_Wiita1980} (hereafter PW80) that provides very satisfactory results. Numerous groups of researcher extensively investigated the physical properties of the astrophysical flows around non-rotating black holes using PW80 potential \citep[and references therein]{Abramowicz_Zurek1981,Matsumoto-elal84,Czerny-Elvis87,Chakrabarti1989,Yang_Kafatos1995,Chakrabarti_Titarchuk1995,Chakrabarti1996,Manmoto-etal97,Narayan-etal97,Hawley-Krolik01,Fukue-Hirai01,Becker-Le03,Das2007,Sarkar_Das2016,Dihingia_etal2018a}

However, in reality, the presumption of non-rotating black hole possibly is too simplistic in the sense that all the cosmological objects are expected to be rotating. 
Hence, the use of Kerr geometry as an appropriate background seems to be inevitable which in general plays key role in studying the accretion phenomenon around rotating black holes. However, solving the general relativistic (GR) hydrodynamic equations around a Kerr black hole is complicated and challenging. To overcome this, various pseudo-potentials for Kerr black hole were proposed \citep{Chakrabarti_Khanna1992,Artemova_etal1996,Semerak_Karas1999,Mukhopadhyay02,Ivanov_Prodanov2005,Chakrabarti_Mondal2006}. All these potentials were prescribed based on certain physical constraints and the problem under considerations. Naturally, all of them have their own limitations and therefore, regime of validity of these potentials are very much restricted. For example, the potential proposed by  \cite{Chakrabarti_Mondal2006} describes the space-time geometry satisfactorily for black hole spin parameter $a_{\rm k} < 0.8$ and thus, this potential can not be used to study the systems which are believed to harbor rapidly rotating black holes like Sgr A*, Cyg X-1, LMC X-1,M33 X-7, 4U 1543-47,  GRO J1655-40, GX 339-4 etc.  \citep{Shafee-etal06,Gou_etal2009,Aschenbach2010,Liu_etal2010,Gou_etal2011,Ludlam-etal15}. Therefore, it is of prime importance in the astrophysical context to ascertain the form of the effective potential corresponding to a Kerr black hole, which will be free from any a priori restrictions mentioned above.

With the increasing sophistication of observational techniques and precession measurements, it would always be prudent to understand any physical phenomena in a model independent way. For example, in the weak gravity regime, Einstein's theory has been proven to be very successful through a large number of observations. Although the high precession measurement in strong gravity regime is yet to be substantiated, however, it has the potential to differentiate new physics beyond Einstein, if any. Keeping this in mind, our goal in this paper will be multi fold. Firstly, to make our analysis model independent, we will be considering a generic axisymmetric space-time, and then formulate the full general relativistic hydrodynamics. While studying different component equations of the hydrodynamics, it turns out that in the static limit, the radial flow equation can be cast into a Newtonian like flow equation. Therefore, the most powerful result we obtain through the present analysis is an analytic expression for the effective potential for any generic axisymmetric black hole space-time. Next, we consider the Kerr metric as a representation of rotating black hole and study the accretion dynamics in detail. In continuation, we infer the limitations of the conventional Newtonian approach while examining the accretion flow around the black holes in the non-relativistic limit using effective potentials. To this end, we mention some of the recent important theoretical developments in the non-relativistic hydrodynamics as a special limit of relativistic hydrodynamics for certain conformal field theory. In the slow fluid velocity limit ($i.e.$, $v/c \ll1$), when the fluid pressure is redefined in a way that 
the thermal motion of the fluid constituents does not violate the above speed limit, the relativistic hydrodynamic equations for a conformal field theory boil down to the incompressible non-relativistic Navier-Stokes equation  \citep{Fouxon_Oz2008}. Soon after, \citet{Bhattacharyya_etal2009} reported similar findings  
by appropriately scaling all fluid and thermodynamic variables, respectively.

In this work, we are also interested to the similar non-relativistic limit. In fact, our goal is to move even further where we quantitatively compare the results obtained from different limits, such as relativistic (R) and non-relativistic (NR; $v/c \ll 1$ and $k_BT/(m_ec^2)\ll 1$) specifically in the context of accretion flow dynamic. Most important findings we observe here is that the conventional Newtonian approach to study the accretion flow around black hole endures inherent limitation that originates due to the adopted deceiving dynamics of the flow in the vicinity of the black hole horizon. Moreover, we confer the essence of these differences exclusively  focusing the relativistic effect on the flow dynamics.

For simplicity, we consider an adiabatic advective accretion flow to obtain the effective potential for a rotating black hole. The conservation equations that govern the dynamics of the accretion flow around rotating black hole are the mass conservation equation, radial momentum conservation equation and the entropy generation equation, respectively. By suitably defining the radial three-velocity ($v$) in co-rotating frame, the radial momentum equation is expressed as the addition of three terms, namely kinetic energy, thermal energy and gravitation energy, respectively at per with the Newtonian flow equation although all the conserved equations under consideration are fully relativistic in nature. With this, we successfully identify the analytic expression of the effective potential in a generic axisymmetric space-time.

In view of the importance of effective potential, we intend to investigate the behavior of accretion flow around a rotating black hole. We find the global transonic solutions that connect the black hole horizon and the outer edge of the disc (equivalently large distance away from the black hole). In reality, during accretion, rotating matter is piled up in the vicinity of the black hole due to the centrifugal repulsion against gravity that eventually triggers the discontinuous transition of the flow variables in the form of shock wave \citep{Fukue1987,Chakrabarti1989,Yang_Kafatos1995,Lu-etal99,Becker_Kazanas2001,Fukumura-Tusuruta04,Das2007,Kumar-Chattopadhyay2014,Sarkar_Das2016,Dihingia_etal2018a,Dihingia_etal2018b}. We calculate the global transonic accretion solutions including shock waves and compare it for all the limiting conditions considering non-rotating, weakly rotating and rapidly rotating black holes. Further, we separate the domain of the parameter space in angular momentum and energy ($\lambda-{\cal E}$) plane according to the nature of flow solutions. We also identify the effective region of the parameter space for a wide range of black hole spin values that admits shock induced global accretion solutions. In this work, we ignore the dissipative processes, namely viscosity, radiative cooling, magnetic fields to avoid complexity. We plan to consider these physical processes in the future study.

In \S 2, we discuss the mathematical background. In \S 3, we describe the governing equations and carried out the critical point analysis. In \S 4, we discuss the global accretion solutions with and without shocks and also classify the shock parameter space.
Finally, in \S 5, we present concluding remarks.

\section{Relativistic Hydrodynamics in general stationary axisymmetric space-time}

As emphasized earlier, we analyze the relativistic hydrodynamic 
equations in a generic stationary axisymmetric space-time. Defining 
property of a static axisymmetic space-time is the existence of two 
commuting killing vectors which we will take along $(t,\phi)$ direction. 
The rest of the space-like coordinates identified as $(r, \theta)$ will 
be assumed to be mutually orthogonal as well as orthogonal to the 
two killing vector fields at each point in the space-time. Therefore, 
with the above choice of coordinate system a generic stationary 
axisymmetric space-time can be expressed as 
\bea 
ds^2& = &g_{\mu\nu} dx^\mu dx^\nu \\
&= & g_{tt}dt^2 + 2g_{t\phi}dtd\phi+ g_{\phi\phi} d\phi^2 + 
g_{rr} dr^2 + g_{\theta\theta} d\theta^2 \nno ,
\label{eq1}
\eea
where $\mu$ and $\nu$ are indices run from $0$ to $3$. Assuming a 
black hole to be located at the center, horizon is identified as 
$g^{rr} =1/g_{rr} =0$. Because of the two killing vectors
$(l_t = \partial_t, l_{\phi} = \partial_{\phi})$, all the metric coefficients
will in general be function of $(r,\theta)$. 

\subsection{Hydrodynamics}
Hydrodynamics is a model independent approach towards the 
understanding of the low energy dynamics of any generic field
theory. The construction is based on the underlying symmetry 
and the associated conservation laws of the theory. If we consider
a Lorentz invariant theory with a global $U(1)$ symmetry, properties
of hydrodynamic flow are studied using the following two conservation
equations for the energy-momentum and particle number as, 
$$
T^{\mu\nu}_{;\nu}=0 ~~{\rm and}~~  j^\mu_{;\mu}=0.
\eqno(2)
$$
Here, the energy momentum tensor $T^{\mu\nu}$ and the particle
number current $j^{\mu}$ are expressed in terms of systematic
derivative expansion of the fluid degrees of freedom consisting
of local energy density $e(r)$, pressure $p(r)$, and the four velocity
$u^{\mu}$ supplemented by the condition $u^{\mu}u_{\mu} =-1$. In 
general, one writes
$$
T^{\mu\nu} = T^{\mu\nu}_{0} + \pi^{\mu\nu} ~~{\rm and}~~ j^{\mu}=j^{\mu}_0 + \pi^{\mu} .
\eqno(3)
$$
These equations (3) are called constitutive relations. The first term
in the right hand side of both equations are zeroth order, and the second
terms contain all derivative terms. For example, the dissipative term 
which contains the first order derivative in fluid velocity, will appear 
in the second term. For the present analysis, we confine ourself 
only to the zeroth order term. Therefore, zeroth order expansion of 
the energy momentum tensor and the four current are written as
$$
T_0 ^{\mu\nu} = (e+p)u^\mu u^\nu + pg^{\mu\nu} ~~{\rm and}~~ j^{\mu}_0 = \rho u^{\mu},
\eqno(4)
$$
where $e$, $p$, and $\rho$ are the local energy density, local isotropic 
pressure and mass density of the flow. Therefore, the final zeroth order 
hydrodynamic equations are given by,
$$
{T_0^{\mu\nu}}_{;\nu}=0 ~~{\rm and} ~~ (\rho u^\nu)_{;\nu}=0.
\eqno(5)
$$

With respect to the fluid flow, we construct projection 
operator $h^i_\mu = \delta^i_\mu + u^i u_\mu$, with  `$i$' takes 
$(1,2,3)$ values. It also satisfies $h^i_{\mu}u^{\mu}=0$. This condition helps us to project the Navier-Stokes equation into three vector equations as
$$
h^i_\mu {T_0^{\mu\nu}}_{;\nu}=(e+p)u^\nu u^i_{;\nu} + (g^{i\nu} + u^i u^\nu)p_{,\nu}=0, 
\eqno(6)
$$
and a scalar equation which is essentially identified as second low of thermodynamics,
$$
u_\mu T^{\mu\nu}_{;\nu}=u^\mu\bigg[\left(\frac{e+p}{\rho}\right)\rho_{,\mu} - e_{,\mu}\bigg]=0.
\eqno(7)
$$
In this work, our goal is to cast the relativistic radial momentum flow equation at per with the 
into Newtonian like. Therefore, we define the following variables in their 
appropriate form: the angular velocity variable 
$v_\phi^2 = (u^\phi u_\phi)/(-u^t u_t)$, and the associated bulk 
azimuthal Lorentz factor as $\gamma_\phi^2=1/(1-v_\phi^2)$. Subsequently, the polar three velocity
is defined as $v_\theta^2 = \gamma_\phi^2(u^\theta u_\theta)/(-u^tu_t)$ and the
associated bulk polar Lorentz factor as $\gamma_\theta^2=1/(1-v_\theta^2)$.
Similarly, the radial three velocity in the co-rotating frame is defined 
as $v^2 = \gamma_\phi^2\gamma_\theta^2v_r^2$, where $v_r^2 = (u^ru_r)/(-u^tu_t)$ and the associated bulk radial Lorentz factor 
$\gamma^2_v = 1/(1-v^2)$. 
Employing these definitions of the velocities in equation (6), we obtain the 
equations corresponding to $i=r$ and $i=\theta$ which are given by,
$$
\gamma^2_v v\frac{\partial v}{\partial r} + \gamma^2_v\gamma_\theta v_\theta\sqrt{\frac{g_{rr}}{g_{\theta\theta}}}\frac{\partial v}{\partial \theta} 
+ \frac{\gamma_\theta v v_\theta}{2\sqrt{g_{rr}g_{\theta\theta}}}\frac{\partial g_{rr}}{\partial \theta} 
- \frac{v_\theta^2\gamma_\theta^2}{2g_{\theta\theta}}\frac{\partial g_{\theta\theta}}{\partial r}
$$
$$
+\frac{1}{e+p}\frac{\partial p}{\partial r}+ \frac{\gamma_\theta v v_\theta}{e+p} \sqrt{\frac{g_{rr}}{g_{\theta\theta}}}\frac{\partial p}{\partial \theta} + \gamma_\theta^2\frac{\partial \Phi^{\rm eff}}{\partial r}=0,
\eqno(8)
$$
and
$$
\gamma_\theta^3 v\frac{\partial v_\theta}{\partial r} + \gamma_\theta^4v_\theta \sqrt{\frac{g_{rr}}{g_{\theta\theta}}}\frac{\partial v_\theta}{\partial \theta} + \gamma^2_v\gamma_\theta v^2v_\theta\frac{\partial v}{\partial r}
$$$$
+ \gamma^2_v\gamma_\theta^2 v v_\theta^2\sqrt{\frac{g_{rr}}{g_{\theta\theta}}}\frac{\partial v}{\partial \theta} - \frac{v}{2}\sqrt{\frac{g_{rr}}{g_{\theta\theta}}}\left(\frac{v}{g_{rr}} \frac{\partial g_{rr}}{\partial \theta} - \frac{\gamma_\theta v_\theta}{\sqrt{g_{rr}g_{\theta\theta}}} \frac{\partial g_{\theta\theta}}{\partial r}\right)
$$$$
+ \frac{\gamma_\theta^2-v^2}{e+p}\sqrt{\frac{g_{rr}}{g_{\theta\theta}}}\frac{\partial p}{\partial \theta} + \frac{\gamma_\theta v v_\theta}{e+p}\frac{\partial p}{\partial r} + \gamma_\theta^2 \sqrt{\frac{g_{rr}}{g_{\theta\theta}}}\frac{\partial \Phi^{\rm eff}}{\partial \theta}=0.
\eqno(9)
$$
As already mentioned, the defining property of a general static, axisymmetry space-time is the existence of two commuting Killing vector fields 
$l^\mu_t\equiv\partial_t$ and $l^\mu_\phi\equiv\partial_\phi$, associated with time translation and azimuthal rotation, respectively. For each globally defined killing vector $l$ there exists an associated conserved quantity say $Q_l$. By using the equations for the mass and energy momentum conservations, in general for a non-dissipative fluid, one can express the aforementioned conserved quantity $Q_{l} = l^{\mu} h u_{\mu}$, which satisfies the conservation equation ${\pounds}_{u}Q_l=0$, where ${\pounds}_{u}$ is Lie derivative along the flow vector $u$. 
Hence, the fluid flow around a static, axisymmetric background leads to the following two conserved quantities,
$$
hu_\phi = {\cal L}~({\rm constant}) \qquad {\rm and} \qquad hu_t= {\cal E}~({\rm constant}),
\eqno(10)
$$
where $h~[=(e+p)/\rho]$ is the enthalpy of the flow and ${\cal E}$  is the 
relativistic Bernoulli constant.
Here, $u_t^2 = \gamma^2/(g^{t\phi}\lambda - g^{tt})$ where $\lambda =-u_\phi/u_t$ is the conserved specific angular momentum of the fluid and $\gamma = \gamma_\phi\gamma_v\gamma_\theta$ is the total bulk Lorentz factor. It is to be noted that equations
(8) and (9) exactly reduce to the Euler equations of the Newtonian 
hydrodynamics (follow \cite{Rezzolla_Zanotti2013}). Thus, these two equations  
describe the relativistic momentum of the flow along radial ($r$) and polar ($\theta$) directions 
where $\Phi^{\rm eff}$ denotes effective pseudo potential and is given by,
$$
\Phi^{\rm eff} = 1+0.5 \ln (\Phi),
\eqno(11a)
$$
where
$$
\Phi =  \frac{g_{t\phi}^2-g_{tt}g_{\phi\phi}}{g_{\phi\phi}+2 \lambda g_{t\phi}+\lambda^2 g_{tt}}.
\eqno(11b)
$$
In the next section, we consider a specific black hole background of astrophysical interest and discuss its consequence on the accretion flow dynamics in details.

\section{Governing equations}
\subsection{Equations describing relativistic accretion flow around Kerr black hole}

In the present paper, we consider a specific stationary axisymmetric space-time for Kerr black hole.
In terms of Boyer-Lindquist coordinates, the components of Kerr metric are expressed as follows  
\citep{Boyer_Lindquist1967},
\bea
g_{tt}& =& -\left(1 - \frac {r r_g}{\Sigma}\right) ~~;~~g_{t\phi} = -\frac{a_{\rm k} r r_g\sin^2\theta}{\Sigma} \nno\\
g_{rr}&=&\frac{\Sigma}{\Delta} ~~;~~ g_{\theta\theta}=\Sigma~~;~~g_{\phi\phi}=\frac{A\sin^2\theta}{\Sigma},\nno
\eea
$$\eqno(12)$$
where $A=(r^2 + a_{\rm k}^2)^2 - \Delta a_{\rm k}^2\sin^2\theta$, 
$\Sigma = r^2 + a_{\rm k}^2\cos^2\theta$ and $\Delta = r^2 - r_g r + a_{\rm k}^2$,
respectively. This metric successfully describes the space-time geometry around a 
rotating black hole of mass $M_{\rm BH}$ and angular momentum $J$. We write the
specific spin of the black hole as $a_{\rm k}=J/ M_{\rm BH}$.   
For convenience, we use a unit system as $G=M_{\rm BH}=c=1$, where $G$ and $c$ are
the gravitational constant and speed of light. Therefore, measurement of speed, mass, 
length, time, angular momentum and energy will be expressed in unit of $c$, 
$M_{\rm BH}$, $GM_{\rm BH}/c^2$, $GM_{\rm BH}/c^3$, $GM_{\rm BH}/c$ and
$M_{\rm BH}c^2$, respectively. It is to be noted that in equations (12), 
$r_g$ refers the Schwarzchild radius and is given by $r_g = 2G M_{\rm BH}/c^2$. 
In this unit system, the effective potential around a Kerr black hole is computed as,
$$
\Phi^{\rm eff}= 1+\ln\bigg[\frac{ A (2 r- \Sigma )\sin ^2\theta-4 a_{\rm k}^2 r^2 \sin ^4\theta}{\Sigma  \lambda  \left(\lambda  \Sigma +4 a_{\rm k} r \sin ^2\theta -2 \lambda  r\right)-A\Sigma\sin ^2\theta }\bigg].
\eqno(13)
$$
In this work, our goal is to solve the hydrodynamic equations around Kerr black hole.
In order to proceed further, we consider a geometrically thin accretion disc which is
confined around the black hole equatorial plane. Therefore, for simplicity, we choose 
$\theta=\pi/2$. Accordingly, the flow motion along the transverse direction is considered to be negligible $i.e.$, $v_\theta = 0$. With this, we have $\gamma_\theta = 1$ and $u_t = \gamma_v\sqrt{\frac{r \Delta}{a_{\rm k}^2 (r+2)-4 a_{\rm k} \lambda +r^3-\lambda ^2 (r-2)}}$.  Moreover, we also neglect
the $\theta$ variation of all the flow variables.
With these approximations, the radial component of equation (8) turns out
to be the well known Navier-Stokes equation, which is given by, 
$$
v\gamma_v^2\frac{dv}{dr} + \frac{1}{h\rho}\frac{dp}{dr} + \frac{d\Phi^{\rm eff}_e}{dr}=0,
\eqno(14)
$$
where $\Phi^{\rm eff}_e$ represents the effective pseudo potential calculated at the equatorial plane ($\theta=\pi/2$) and is given by,
$$
\Phi^{\rm eff}_e =1+\frac{1}{2} \ln\bigg[\frac{r\Delta}{a_{\rm k}^2 (r+2)-4 a_{\rm k} \lambda +r^3-\lambda ^2 (r-2)}\bigg].
\eqno(15)
$$

Similarly, the entropy generation equation is calculated from equation (7) as,
$$
\left(\frac{e+p}{\rho}\right)\frac{d\rho}{dr} - \frac{de}{dr}=0.
\eqno(16)
$$
The second part of equations (5) which is basically the continuity equation, is rewritten in an integrated form as, 
$$
\dot{M} =-4\pi v \gamma_v \rho H \sqrt{\Delta},
\eqno(17)
$$
where $\dot{M}$ is the accretion rate and $H$ is the 
local half-thickness of the disc and its functional
form under thin disc approximation is computed as  \citep{Riffert_Herold1995,Peitz_Appl1997},
$$
H^2 = \frac{pr^3}{\rho \mathcal{F}}~,~
\mathcal{F}=\gamma_\phi^2\frac{(r^2 + a_{\rm k}^2)^2 + 2\Delta a_{\rm k}^2}
{(r^2 + a_{\rm k}^2)^2 - 2\Delta a_{\rm k}^2}.
\eqno(18)
$$

In order to solve equations (14), (16) and (17), one needs to consider a relation among $e$, $\rho$ and $p$, commonly known as equation of state (EoS).
In the subsequent analysis, we adopt an EoS proposed by \cite{Chattopadhyay-Ryu2009} that 
agrees quite satisfactorily 
with the exact EoS of the fluid \citep{Chandrasekhar1939,Synge1957,Cox_Giuli1968}.
For a fluid consisting of electrons, positrons and ions, the EoS is given by,
$$
e = n_em_ef=\frac{\rho}{\tau}f,
\eqno(19)
$$
where $\rho = n_em_e\tau$ and $\tau = [2 - \xi(1 - 1/\chi)]$. Here, $n_e$ ($n_p$)
and $m_e$ ($m_p$) represent the number density and mass of the electron (ion). 
Moreover, we define $\xi = n_p/n_e$ and $\chi=m_e/m_p$, respectively. 
Throughout our study, we use $\xi=1$, until otherwise stated. Finally, the
functional form of $f$ is given by,
$$
f = (2-\xi)\bigg[1 + \Theta\left(\frac{9\Theta + 3}{3\Theta + 2}\right)\bigg] +
\xi\bigg[ \frac{1}{\chi} + \Theta\left(\frac{9\Theta + 3/\chi}
{3\Theta + 2/\chi}\right)\bigg] ,
\eqno(20)
$$
where we define the dimensionless temperature of the fluid as $\Theta = k_BT/m_ec^2$.
In addition, polytropic index $(N)$, specific heat ratio $(\Gamma)$ and sound 
speed $(C_s)$ are define as,
$$
N = \frac{1}{2}\frac{df}{d\Theta}; \quad \Gamma = 1 + \frac{1}{N}~~{\rm and}
\quad C_s^2 = \frac{\Gamma p}{e+p} = \frac{2\Gamma\Theta}{f + 2\Theta}.
\eqno(21)
$$

After some algebraic steps involving equations (14), (16), (17) and (19), we calculate the wind equation as,
$$
\frac{dv}{dr}= \frac{\mathcal{N_{\rm R}}}{\mathcal{D}_{\rm R}},
\eqno(22)
$$
where denominator $\mathcal{D}_{\rm R}$ is given by,
$$
\mathcal{D}_{\rm R} = \gamma_v^2\bigg[v- \frac{2C_s^2}{v(\Gamma +1)}\bigg],
\eqno(23)
$$
and numerator $\mathcal{N}_{\rm R}$ is given by,
$$
\mathcal{N}_{\rm R}=\frac{2C_s^2}{\Gamma + 1}\bigg[ \frac{\left(r-a_{\rm k}^2\right)}{r\Delta} + \frac{5}{2r} - \frac{1}{2\mathcal{F}}\frac{d\mathcal{F}}{dr}\bigg] - \frac{d\Phi^{\rm eff}_e}{dr}.
\eqno(24)
$$

Similarly, the gradient of the temperature is obtain as,
$$
\frac{d\Theta}{dr}=-\frac{2\Theta}{2N + 1}\bigg[\frac{\left(r-a_{\rm k}^2\right)}{r\Delta}
+\frac{\gamma_v^2}{v}\frac{dv}{dr} + \frac{5}{2r} - \frac{1}{2\mathcal{F}}\frac{d\mathcal{F}}{dr}\bigg].
\eqno(25)
$$
In equations (24) and (25), the logarithmic derivatives of $\mathcal{F}$ is calculated as,
$$
\frac{1}{\mathcal{F}}\frac{d\mathcal{F}}{dr} = \gamma_\phi^2\lambda\Omega' + 
4 a_{\rm k}^2 \left(a_{\rm k}^2+r^2\right)\frac{ \left(a_{\rm k}^2+r^2\right) \Delta {'}
	-4 r \Delta}{\left(a_{\rm k}^2+r^2\right)^4-4 a_{\rm k}^4 \Delta^2},
\eqno(26)
$$
where $\Delta'=2(r-1)$ and
$$
\Omega' = - 2\frac{a_{\rm k}^3-2 a_{\rm k}^2 \lambda +a_{\rm k} \left(\lambda ^2+3 r^2\right)+\lambda  (r-3) r^2}{\left(a_{\rm k}^2 (r+2)-2 a_{\rm k} \lambda +r^3\right)^2}.
\eqno(27)
$$
It is to be noted that the ratio of the radial flow velocity ($v$) to the speed of light ($c$) always remain $v/c \lesssim 0.1$ even in the region $r > 4r_g$ \citep[and references therein]{Sarkar-etal2018}. Therefore, for all practical purpose, we can safely set $\gamma_v \rightarrow 1$ and hence, the radial momentum equation (14) reduces into the simplified form as,
$$
v \frac{dv}{dr} + \frac{1}{h \rho}\frac{dp}{dr} + \frac{d\Phi^{\rm eff}_e}{dr}=0.
\eqno(28)
$$
However, it would be worthy to compare results obtained separately using equation (14) and equation (28) which will be discussed in the subsequent sections. For convenience, we refer the analysis that incorporates equation (28) as semi-relativistic (SR) limit.

\subsection{Equations in non-relativistic limit}

A non-relativistic accretion flow is characterized by $v\ll1$ all throughout. Therefore, in this limit, the Lorentz factor becomes $\gamma_v = 1$. Moreover, one also needs to maintain the temperature and pressure of the fluid, so that thermal speed should not exceed the non-relativistic limit ($i.e.$, $\Theta \ll1$). With this consideration, the enthalpy of the flow becomes
$h(r) \sim 1 $ and hence, 
$(h\rho)^{-1}(dp/dr)\sim {\rho}^{-1}(dp/dr)$.  
With this, the radial momentum equation is reduced as,
$$
v\frac{dv}{dr} + \frac{1}{\rho}\frac{dp}{dr} + \frac{d\Phi^{\rm eff}_e}{dr}=0.
\eqno(29)
$$ 
It may be noted that equation (29) is the well known Euler equation in Newtonian
hydrodynamics. Upon integrating equation (29), we obtain the specific energy 
(including the rest mass energy) of the flow as,
$$
{\cal E}_{\rm NR}= \frac{v^2}{2} + h + \Phi^{\rm eff}_e - 1,
\eqno(30)
$$
where we use the relation ${\rho}^{-1}(dp/dr)=dh/dr$. Now, it is clear that in the non-relativistic
limit, the radial momentum equation transforms into the Newtonian hydrodynamics 
equations with an effective potential $\Phi^{\rm eff}_e$. 
In the limit $r\gg2$, equation (15) reduces to the Newtonian effective potential
experienced by a particle around a Newtonian object and is given by,
$$
\Phi^{\rm eff}_e\bigg|_{r\gg2} =\Phi_{\rm Newton}= 1 - \frac{1}{r} + \frac{\lambda^2}{2r^2}.
$$
Needless to mention that the entropy generation equation (equation 16) and the mass conservation equation (equation 17) remain unaltered in the non-relativistic domain. Using equations (16), (17) and (29), we again calculate the wind equation which is given by,
$$
\frac{dv}{dr}=\frac{\cal{N}_{\rm NR}}{\mathcal{D_{\rm NR}}},
\eqno(31)
$$
where denominator $\mathcal{D_{\rm NR}}$ is given by,
$$
\mathcal{D_{\rm NR}} = \bigg[v- \frac{2C_s^2h}{v(\Gamma +1)}\bigg],
\eqno(32)
$$
and numerator $\mathcal{N_{\rm NR}}$ is given by,
$$
\mathcal{N_{\rm NR}}=\frac{2C_s^2h}{(\Gamma + 1)}\bigg[ \frac{\left(r-a_{\rm
		k}^2\right)}{r\Delta} + \frac{5}{2r} -
\frac{1}{2\mathcal{F}}\frac{d\mathcal{F}}{dr}\bigg] - \frac{d\Phi^{\rm eff}_e}{dr}.
\eqno(33)$$
Here, subscript `NR' denotes the quantities calculated considering non-relativistic approximation.

The gradient of temperature is computed as,
$$
\frac{d\Theta}{dr}=-\frac{2\Theta}{2N + 1}\bigg[\frac{\left(r-a_{\rm k}^2\right)}{r\Delta}
+\frac{1}{v}\frac{dv}{dr} + \frac{5}{2r} - \frac{1}{2\mathcal{F}}\frac{d\mathcal{F}}{dr}\bigg].
\eqno(34)$$ 

In the subsequent sections, we carry out the comparative analysis considering relativistic, semi-relativistic and non-relativistic equations, and 
show how the flow properties obtained from non-relativistic hydrodynamics significantly
deviate from that computed from relativistic dynamics specifically near black hole horizon.

\subsection{Critical point analysis}
During the course of accretion around black hole, flow starts to move 
inwards sub-sonically from the outer edge of the disc and eventually enters in to
the black hole with supersonic speed. Since the flow accretes smoothly along the
streamline, the radial velocity gradient remains real and finite always. However, 
equations (23) and (32) indicate that denominator (${\cal D}_{\rm R}$ and
${\cal D}_{\rm NR}$) of the wind equations may vanish at some radial coordinate. 
To maintain the smoothness of the flow, numerator (${\cal N}_{\rm R}$ and
${\cal N}_{\rm NR}$) of the wind equations must also go to zero there. Such a
special point where the gradient of radial velocity take the form as
$(dv/dr)_{\rm c} \rightarrow 0/0$ is called as critical point ($r_{c}$). Setting numerator and 
denominator simultaneously equal to zero, we obtain the critical point conditions
which are given below for both relativistic and non-relativistic cases.

\subsubsection{Critical Point Conditions for Relativistic flow}

For relativistic flow, setting ${\cal D}_{\rm R}=0$ in equation (23), we obtain the
radial velocity ($v_{\rm c}$) at the critical point ($r_{\rm c}$) as,
$$
v_{\rm c}^2= \frac{2C_{s{\rm c}}^2}{(\Gamma_{\rm c} +1)}.
\eqno(35)
$$

Further, setting ${\cal N}_{\rm R}=0$ in equation (24),
we get the sound speed ($C_{s\rm c}$) at $r_{\rm c}$ as,
$$
C_{s\rm c}^2 = \frac{\Gamma_{\rm c}+1}{2}\left( \frac{d\Phi^{\rm eff}_e}{dr}\right)_{\rm c}
\left[ \frac{\left(r_{\rm c}-a_{\rm k}^2\right)}{r_{\rm c}\Delta_{\rm c}} +
\frac{5}{2r_{\rm c}} - \frac{1}{2\mathcal{F}_{\rm
		c}}\frac{d\mathcal{F}_{\rm c}}{dr}\right]^{-1}.
\eqno(36)
$$

\subsubsection{Critical Point Conditions for Non-relativistic flow}
For non-relativistic flow, setting ${\cal D}_{\rm NR}=0$ in equation (32), we calculate 
the radial velocity ($v_{\rm c}$) at $r_{\rm c}$ as,
$$
v_{\rm c}^2= \frac{2C_{s{\rm c}}^2h_{\rm c}}{(\Gamma_{\rm c} +1)}.
\eqno(37)
$$

As before,  we set ${\cal N}_{\rm NR}=0$ in equation (33) to get the 
sound speed at the critical point as,
$$
C_{s\rm c}^2 = \frac{\Gamma_{\rm c}+1}{2h_{\rm c}}\left(\frac{d\Phi^{\rm eff}_e}{dr}\right)_{\rm c}
\left[ \frac{\left(r_{\rm c}-a_{\rm k}^2\right)}{r_{\rm c}\Delta_{\rm c}} +
\frac{5}{2r_{\rm c}} - \frac{1}{2\mathcal{F}_{\rm
		c}}\frac{d\mathcal{F}_{\rm c}}{dr}\right]^{-1}.
\eqno(38)
$$
In the above, subscript `${\rm c}$' refers the flow variables at $r_{\rm c}$.
Since the gradient of the radial velocity takes the `0/0' form at $r_{\rm c}$, we apply l'Hospital rule to calculate
$dv/dr|_{\rm c}$ at $r_{\rm c}$, which is given by,
$$
\frac{dv}{dr}\bigg|_{\rm c} = \frac{-B\pm\sqrt{B^2-4AC}}{2A}.
\eqno(39)
$$
In equation (39), $A$, $B$ and $C$ are functions of flow variables and their
explicit expressions are given in the Appendix. As it is already pointed out that accretion
solutions around black hole must be transonic in nature and therefore, flow must
contain at least one critical point \citep{Liang_Thompson1980,Abramowicz_Zurek1981}.
Depending on the input parameters, accretion flow may possess multiple critical
points as well. When both values of $(dv/dr)_{\rm c}$ are real with opposite sign, the
critical point is called as saddle type and when $(dv/dr)_{\rm c}$ becomes imaginary,
the point is called as `O' type critical point. It may be noted that when $(dv/dr)_{\rm c}$ is negative, it corresponds to accretion solution and the positive $(dv/dr)_{\rm c}$ yields the wind solution. In this work, we are interested to accretion solutions only and therefore, we keep the wind solutions aside for future study.

\section {Results}

\subsection{Computation of Critical Points}

The procedure to calculate the critical point location in all kinds of flows under consideration is identical and hence, we present the methodology for relativistic flow only. For a given set of input parameters, namely ${\cal E}$, $\lambda$ and $a_{\rm k}$, we calculate the critical point location by solving equations (10), (20), (35) and (36) simultaneously. Since any 
realistic accretion flow passes through the saddle type critical point
only \citep{Chakrabarti-Das04,Das2007}, in this study, 
we focus on those accretion solutions that contains saddle type critical points. 
Accordingly, hereafter we refer the saddle type critical point as critical
point in the subsequent analysis. When
flow possesses multiple critical points, one usually forms very close to the
black hole horizon which is called as inner critical point ($r_{\rm in}$) and the other
forms far away form the horizon called as outer critical point ($r_{\rm out}$). 
In this scenario, accretion flow successfully connects the black hole horizon and
the outer edge of the disc, as it either passes through inner or outer
critical point. Interestingly, another viable possibility also exists here.
Rotating inflowing matter when first crosses 
the outer critical point ($r_{\rm out}$) to become supersonic, it experiences centrifugal repulsion
that eventually triggers the centrifugally supported shock transition in the
flow variables \citep[and references therein]{Dihingia_etal2018a} where supersonic
pre-shock flow jumps in to the subsonic branch of the post-shock flow. In the
subsonic branch, flow momentarily slows down, however, gradually gains its radial
velocity due to the influence of strong gravity and finally enters into the black hole
after passing through the inner critical point ($r_{\rm in}$). Solutions of these kinds are 
physically accepted and 
called as the shock induced global accretion solutions around black hole. The position of the shock transition is known as shock location $(r_{s})$ which provides the measure of
the size of post-shock corona (hereafter PSC). In the subsequent sections, we present the 
elaborate discussion on shock solutions.

\begin{figure}
	\includegraphics[scale=0.4]{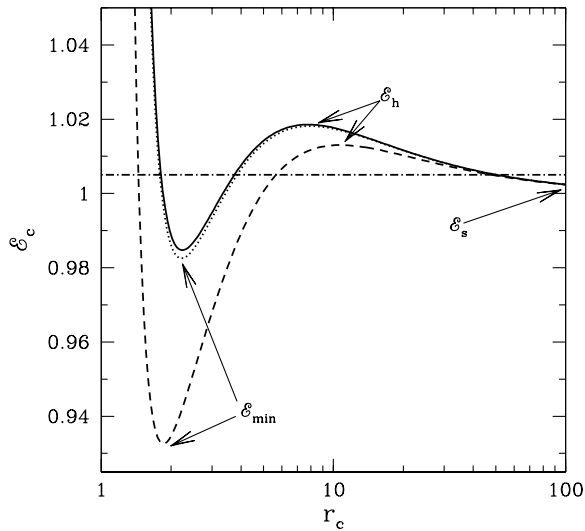}
	\caption{Variation of energy (${\cal E}_{\rm c}$) measured at the critical points ($r_{\rm c}$) as function of $r_{\rm c}$. Solid, dotted and dashed curves represent the results obtained for relativistic (R), semi-relativistic (SR) and non-relativistic (NR) flow, respectively. Here, we choose $\lambda=1.90$, and $a_{\rm k}=0.99$. See text for details.
	}
	\label{fig:fig02}
\end{figure}

In order to understand the transonic nature of the accretion flow, in Fig. 1, we
depict the variation of energy $({\cal E}_{\rm c})$ as function of the critical
point locations ($r_{\rm c}$).
In the figure, the critical points are plotted in logarithmic scale while the energy 
is plotted in linear scale. Solid, dotted and dashed curves represent the
results corresponding to relativistic flow, semi-relativistic flow and 
non-relativistic flow, respectively. Here, we choose $\lambda = 1.90$ and $a_{\rm k} = 0.99$. 
We observe that when
critical points form at a large distance, the flow energy in all the cases remain 
same, however, when critical points form close to the horizon, flow energy 
differs considerably at least for the non-relativistic flow. The small difference
in energy between relativistic and semi-relativistic flows justifies the adopted
approximation that the value of the radial Lorentz factor ($\gamma_v$) deviates only slightly from unity for semi-relativistic flow.
We draw a horizontal dot-dashed line corresponds to ${\cal E}=1.005$ that
intersects with all the curves thrice. This indicates that flow with  
$({\cal E}, \lambda) = (1.005, 1.90)$ possess multiple critical points in all three
cases. In fact, it also indicate that for a flow with fixed $\lambda$, there
is a range ${\cal E}_{\rm s} \le {\cal E} \le {\cal E}_{\rm h}$ for which 
flow possesses three critical points. Needless to mention that both
${\cal E}_{\rm s}$ and ${\cal E}_{\rm h}$, marked in the figure, depend on the $\lambda$
and $a_{\rm k}$, respectively. For ${\cal E}>{\cal E}_{\rm h}$, flow 
possesses only single critical point and for
${\cal E}_{\rm min} < {\cal E} \le{\cal E}_{\rm s}$,
flow possesses two critical points. When ${\cal E} < {\cal E}_{\rm min}$,
critical point ceases to exist.

\begin{figure}
	\includegraphics[scale=0.4]{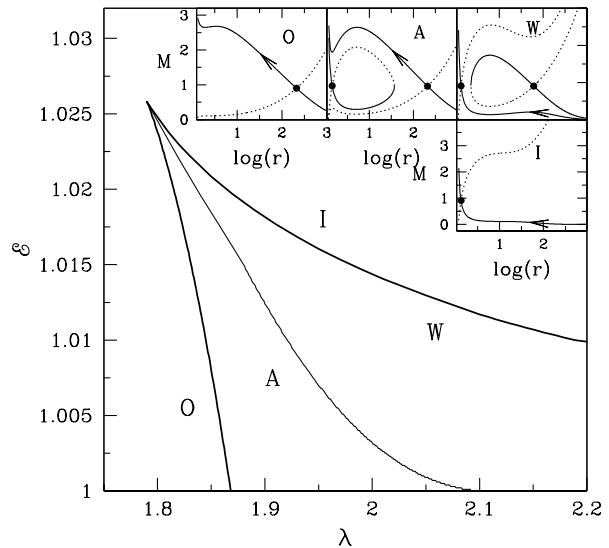}
	\caption{Division of parameter space in $\lambda-{\cal E}$ plane on the basis of flow solutions for semi-relativistic flow. Four regions are marked as `O', `A', `W' and `I' and the corresponding representative solutions (variation of Mach number $M=v/C_s$) are depicted in the boxes. In each box, filled circle represents the location of critical point and arrow indicates the overall direction of the accretion flow motion. See text for details.
	}
	\label{fig:fig01}
\end{figure}

\subsection{Procedure to Compute Global Accretion Solutions}

To obtain a transonic accretion solution, we first calculate the critical point location ($r_{\rm c}$)
for a given set of input parameters (${\cal E}, \lambda, a_{\rm k}$). Afterwards, we employ the critical point conditions (either (35-36) or (37-38)
to calculate the radial velocity and temperature of the flow at the critical point. These
values are used as the initial conditions to integrate the wind equation. In case of relativistic and semi-relativistic
flows, we integrate equation (22), first staring from the critical point up to very close to the
horizon and again from critical point up to a large distance ($x_{\rm edge}$, equivalently the outer
edge of the disc). Eventually, by joining these two parts of the solution, we obtain a global transonic accretion solution around a
rotating black hole. In actuality, one would
get the identical accretion solution provided the integration of the wind equation
is started with the flow variables at $x_{\rm edge}$. For non-relativistic
flows, equations (31) is integrated to obtain the transonic accretion solutions.

\subsection{Parameter Space for Multiple Critical Points}

As already pointed out that depending on the input parameters, an accretion flow may possess multiple critical points. In Fig. 2, we separate the effective domain of the parameter space spanned by ${\cal E}$ and $\lambda$ for global accretion solutions containing multiple critical points. We obtain the result for semi-relativistic flow using $a_{\rm k}=0.99$. Here, we identify four distinctly different regions of the parameter space named as O, A, W and I, based on the type of the solution topologies. In the insets, we display the representative plots of the global solutions which are obtained for the set of input parameters (${\cal E}, \lambda$) chosen from these identified regions of the parameter space as marked in the figure. 
In all the plots, filled circles represent the location of the critical points and arrows indicate the direction of flow motion corresponding to the smooth global accretion solutions. The result corresponding to O-type solution is obtained for (${\cal E}, \lambda) =(1.001, 1.86)$ where outer critical point is located at $r_{\rm out}=211.5867$. We obtain the A-type solution using (${\cal E}, \lambda) =(1.001,2.00)$ and solution of this type contains both inner and outer critical points as $r_{\rm in}=1.4203$ and $r_{\rm out}=210.0059$, respectively. Similarly, for W-type solution, we consider (${\cal E}, \lambda) =(1.004,2.05)$ and obtain $r_{\rm in}=1.3372$ and $r_{\rm out}=60.9931$. Finally, we calculate the I-type solution for (${\cal E}, \lambda) =(1.013,2.05)$ that only passes through the inner critical point at $r_{\rm in}=1.3341$. Note that the regions marked as A and W provide the global accretion solutions that contain multiple critical points.

\begin{figure}
	\includegraphics[scale=0.45]{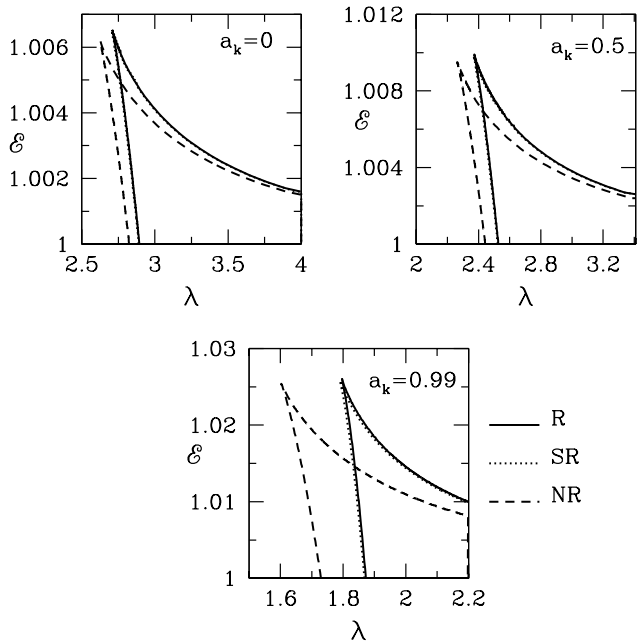}
	\caption{Comparison of parameter space in $\lambda-{\cal E}$ plane that admits the flow to contain multiple critical points. Region bounded by solid, dotted and dashed curves are for relativistic (R), semi-relativistic (SR) and non-relativistic (NR) flow, respectively. Top-left, top-right and bottom panels are for $a_{\rm k} = 0.0, 0.5$ and $0.99$. See text for details.
	}
	\label{fig:fig05}
\end{figure}

Next, we compare the domain of the parameter space for multiple critical points considering the nature of the flow to be relativistic, semi-relativistic and non-relativistic, respectively. The comparative study is carried out around non-rotating ($a_{\rm k}=0$), moderately rotating ($a_{\rm k}=0.5$) and rapidly rotating ($a_{\rm k}=0.99$) black holes and the obtained results are depicted in Fig. 3. In each panel of Fig. 3, effective domain bounded with solid, dotted and dashed curves are obtained for relativistic, semi-relativistic and non-relativistic flows and the values of $a_{\rm k}$ is marked. We observe that parameter spaces for multiple critical points corresponding to relativistic and semi-relativistic cases are in agreement irrespective to black hole spin ($a_{\rm k}$) value. However, the parameter space obtained for non-relativistic flow deviates considerably from the relativistic case and the deviation increases with the increase of $a_{\rm k}$. In reality, as $a_{\rm k}$ is increased, the position of the inner critical points is shifted towards the horizon where space-time is largely distorted and thus the resulting discrepancy is observed.
Overall, the above findings clearly indicate that the non-relativistic approximation bears noticeable limitation as it fails to describe the accretion flow dynamics around rotating black holes satisfactorily.

\subsection{Global Accretion Solution Containing Shock}

It is already anticipated (see \S 4.1) that an accretion flow can pass through the multiple critical points when flow experiences discontinuous shock transition in between them ($i.e.$, $r_{\rm in}$, and $r_{\rm out}$). 
In general, the formation of shock waves is natural in the astrophysical context \citep[and references therein]{Chakrabarti1990,Schaal_etal2015, Schaal_etal2016} due to the fact that shock induced accretion solutions possesses higher entropy content than the shock free solutions \citep{Becker_Kazanas2001}. To compute the shock location, we utilize the relativistic shock conditions which are given by \cite{Taub1948},

$$\begin{aligned}
&[\rho u^r]=0, \qquad [(e+p)u^tu^r]=0,\\
&{\rm and}\quad[(e+p)u^ru^r + pg^{rr}]=0,\\
\end{aligned}\eqno(40)$$
where the difference of quantities across the shock is denoted by the square brackets.

\begin{figure}
	\includegraphics[scale=0.45]{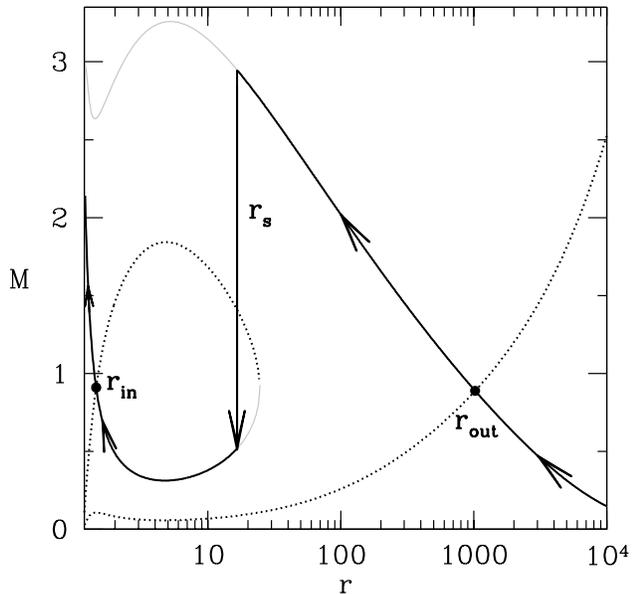}
	\caption{The Mach number ($M=v/C_s$) of the semi-relativistic flow is plotted with radial coordinate ($r$) for ${\cal E} = 1.0001$, $\lambda = 1.989$ and $a_{\rm k} = 0.99$. Thick solid curve represents the accretion solution while the dotted curves refers the wind solution. Vertical arrow denotes location of shock transition ($r_s$) and arrows indicate the overall direction of the accretion flow motion. Filled circles refer the critical points where inner critical point ($r_{\rm in}$) and outer critical point ($r_{\rm out}$) are marked. See text for details.
	}
	\label{fig:fig03}
\end{figure}

In Fig. 4, we illustrate a representative accretion solutions containing multiple critical points where Mach number ($M=v/C_s$) of the flow is plotted as function of radial distance. Here, the flow parameters are chosen as ${\cal E}=1.0001$ and $\lambda=1.989$. Moreover, we consider $a_{\rm k}=0.99$. The solution consists of two parts passing through two critical points. The one passing through the outer critical point truly establishes the connection between the black hole horizon and the outer edge of the disc, whereas the other one passing through the inner critical point is closed and connects the horizon only. In reality, during the course of accretion process, flow first crosses the outer critical point at $r_{\rm out}=1022.5621$ and continues to proceed towards the black hole supersonically. Meanwhile, shock conditions are satisfied and flow experiences discontinuous transition at $r_{s}=16.5533$.
In the figure, solid vertical arrow indicates the location of shock transition where flow jumps from supersonic to subsonic branch. Due to gravity, subsonic flow gains it radial velocity gradually and eventually enters into the black hole after passing through the inner critical point at $r_{\rm in}=1.4446$. It may be noted that accretion flow generally prefers to pass through the shock as the entropy content in the subsonic branch is higher compared to the supersonic branch \citep{Becker_Kazanas2001}. The arrows point the overall motion of the global accretion solution that contains shock wave. In addition, dotted curves through $r_{\rm in}$ and $r_{\rm out}$ represent solution corresponding to wind branch.

\begin{figure}
	\includegraphics[scale=0.45]{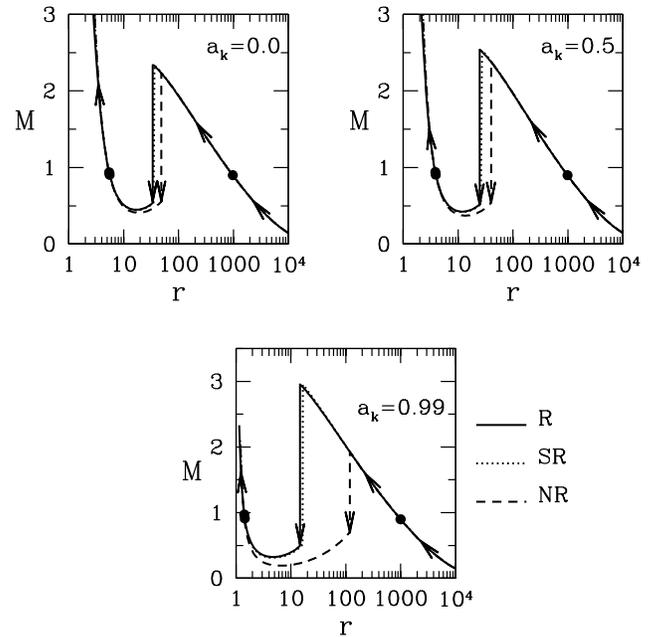}
	\caption{Comparison of shock induced global accretion solutions obtained from relativistic (R, solid), semi-relativistic (SR, dotted) and non-relativistic (NR, dashed) flows. Here, we choose ${\cal E} = 1.0001$ and $\lambda = 3.15$ and $a_{\rm k} = 0.0$ for top-left panel, ${\cal E} = 1.0001$ and $\lambda = 2.75$ and $a_{\rm k} = 0.5$ for top-right panel and ${\cal E} = 1.0001$ and $\lambda = 1.989$ and $a_{\rm k} = 0.99$ for bottom panel. Critical points are shown using filled circles and arrows indicate the direction of flow motion. See text for details.
	}
	\label{fig:fig04}
\end{figure}

In Fig. 5, we compare the shock induced global accretion solutions corresponding to relativistic, semi-relativistic and non-relativistic flows, respectively. The results depicted in top-left, top-right and bottom panels are for non-rotating ($a_{\rm k}=0$), moderately rotating ($a_{\rm k}=0.5$) and rapidly rotating ($a_{\rm k}=0.99$) black holes. Here, we consider the energy of the flow as ${\cal E}=1.0001$ for all cases and choose the angular momentum of the flow as $\lambda= 3.15, 2.75$ and $1.989$ for $a_{\rm k}=0, 0.5$ and $0.99$, respectively. In each panel, solid, dotted and dashed curves represent 
solutions obtained for relativistic, semi-relativistic and non-relativistic flow and sharp vertical arrows indicate the shock positions. Moreover, filled circles denote the critical point locations and arrows indicate the overall direction of the accretion flow starting from the outer edge of the disc up to the horizon. Here also we find that the shock locations computed for relativistic and semi-relativistic flows are in close agreement and this continues even with the increase of $a_{\rm k}$. On the contrary, the obtained shock location for non-relativistic flow differs noticeably from the relativistic solutions and as before, the amount of deviation is increased with $a_{\rm k}$. Quantitative comparison of the transonic and shock properties are given in Table 1.

	\begin{table}[b]
		\caption{Comparison of transonic and shock properties. Here, we choose ${\cal E} = 1.0001$ for all cases.}
		\begin{ruledtabular}
		\begin{tabular}{ccccccc} 
			$a_{\rm k}$&$\lambda$& &$r_{\rm in}$&$r_{\rm out}$&$r_s$& \% Error \\			
			&         & &            &             &     & in $r_s$\\ [0.5ex] 
			\hline \hline 
			0    & 3.15    & GR & 5.5779 & 998.7680 & 33.8730    &  ---  \\
			&              & SR & 5.5674 & 998.5265 & 35.9969    &  6.27 \\
			&              & NR & 5.3942 & 995.6709 & 48.2825    & 42.53 \\
			\hline
			0.5  & 2.75    & GR & 3.9374 & 1008.3744 & 24.6059   &  ---  \\ 
			&              & SR & 3.9284 & 1008.1356 & 26.6320   &  8.23\\
			&              & NR & 3.7751 & 1005.3147 & 39.8949   & 62.13\\ 
			\hline
			0.99 & 1.989   & GR & 1.4485 & 1022.7973 & 14.6729   &  ---   \\ 
			&              & SR & 1.4446 & 1022.5621 & 16.5533   & 12.81\\
			&              & NR & 1.3496 & 1019.7915 & 117.7705  &702.63\\
		\end{tabular}
		\end{ruledtabular}		
	\end{table}

\subsection{Parameter Space for Shock}

\begin{figure}
	\includegraphics[scale=0.45]{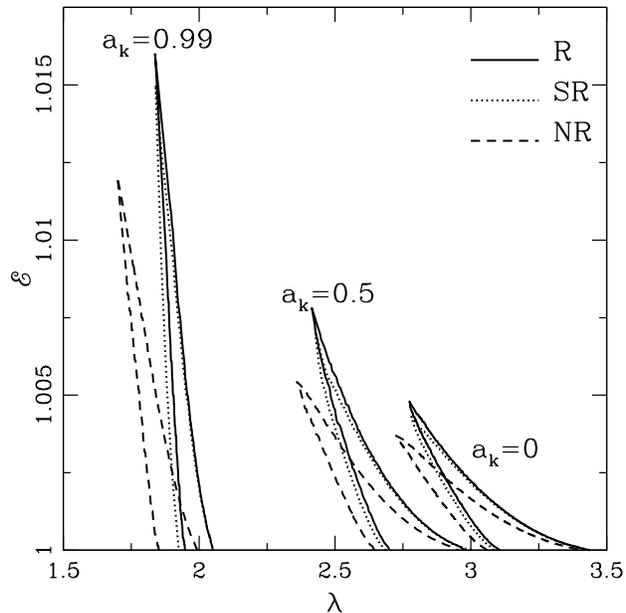}
	\caption{Comparison of shock parameter space in $\lambda-{\cal E}$ plane. Solid, dotted and dashed curves represent the results obtained from relativistic (R), semi-relativistic (SR) and non-relativistic (NR) flows. Here, chosen $a_{\rm k}$ values are marked. See text for details.
	}
	\label{fig:fig06}
\end{figure}

The presence of shock wave in accretion flow seems to plays an important role in determining the black hole spectrum as indicated in \cite{Chakrabarti_Titarchuk1995,Mandal2005}. Due to shock transition, post-shock flow containing hot and dense electrons inverse Comptonizes the soft photons from the cooled pre-shock flow and eventually emerges hard radiations. In addition, electrons are energized while crossing the shock front due to the shock acceleration mechanism and produce non-thermal spectrum. Since shocks are viable and directly involved in deciding the spectral properties
of the black hole sources,
it is therefore worthy to examine whether the shock solutions discussed in the previous section are isolated solutions or not.
For that, 
we continue the study of shock induced global accretion solutions and make an attempt to accomplish the range of flow parameters that admit shocks. The obtained results are displayed in Fig. 6, where we identify the boundary in $\lambda-{\cal E}$ plane that encompasses the effective region of the parameter space for shock around rotating black holes and separate it from the shock free region. 
Here, we compute the shock parameter space considering non-rotating ($a_{\rm k}=0.0$), moderately rotating ($a_{\rm k}=0.5$) and rapidly rotating ($a_{\rm k}=0.99$) black holes and they are marked in the figure. The region bounded with solid, dotted and dashed curves represent the results obtained for 
relativistic, semi-relativistic and non-relativistic flows, respectively. We observe that shock parameter space shifts towards the lower angular momentum and higher energy domain as $a_{\rm k}$ is increased. This happens due to the spin-orbit coupling term in Kerr geometry. 
We notice that the shock parameter spaces of relativistic and semi-relativistic flows are in excellent agreement, but the parameter space computed for non-relativistic flow do show significant deviation from the relativistic result and the deviation increases with $a_{\rm k}$. In particular, for $a_{\rm k}=0.99$, the common overlap of the parameter spaces is seen to be marginal. With this, we argue that non-relativistic approximation for studying the accretion flow dynamics around rotating black hole seems to be incongruous. 

\section{Conclusions}

In this work, we first formulate the set of hydrodynamic equations that describe the accretion flow in a general axisymmetric background and identify an effective potential $\Phi^{\rm eff}$ (see equations (11a,b)). Subsequently, we consider the disc to be confined on the equatorial plane ($i.e.$. $\theta = \pi/2$) and investigate the behavior of relativistic accretion flow around Kerr black hole.
Further, since the radial velocity of the accreting matter in general remains within a few percent of the speed of light even in the vicinity of the horizon ($i.e,~r>4r_g$), we assume $\gamma_v \rightarrow 1$ all throughout the flow. With this consideration, which is named as semi-relativistic limit, we continue to study the accretion flow around rotating black hole. 
It is to be noted that the equations of mass conservation and entropy generation
are not affected by this semi-relativistic approximation.
In addition, we also explore the possibility, where both radial velocity and thermal energy are small as $v \ll 1$ and $h(r) \sim 1$, and this scenario is referred as non-relativistic limit. Finally, we compare the results for the aforementioned three different approaches.  
It is noteworthy that 
the effective potential remains unaltered due to the assumptions adopted in those approaches and it is calculated as
$$
\Phi^{\rm eff}_e =1+\frac{1}{2} \ln\bigg[\frac{r\Delta}{a_{\rm k}^2 (r+2)-4 a_{\rm k} \lambda +r^3-\lambda ^2 (r-2)}\bigg].
$$

Below we summarize our findings based on the present work.

(1) We carry out critical point analysis considering relativistic, semi-relativistic and non-relativistic flows. Excellent agreement is seen between the results obtained from both relativistic and semi-relativistic limit as far as the transonic properties are concerned. However, in the non-relativistic limit, results deviate significantly (see Fig. 1). 

(2) We separate the domain of the parameter space in $\lambda - {\cal E}$ plane based on the nature of solution topologies. We realize that large region of the parameter space permits the existence of multiple critical points which is one of the main criteria to harbor shock wave in accretion flow (see Fig. 2). Moreover, we find that parameter spaces for multiple critical points match sufficiently accurately for relativistic and semi-relativistic flows, but profound difference is seen in the case of non-relativistic flow and the deviation increases with $a_{\rm k}$ (see Fig. 3). 

(3) Considering the semi-relativistic flow, we obtain the shock induced global accretion solution around rapidly rotating black hole (see Fig. 4). Further, we compare the shocked solutions among the relativistic, semi-relativistic and non-relativistic flows having identical outer boundary conditions. We find that the position of shocks in relativistic and semi-relativistic flows agrees well with a deviation of $6-12\%$ for $0\le a_{\rm k}\le 0.99$. But, the difference of shock position between relativistic and non-relativistic flows happens to be very large which becomes monumental ($>62\%$) for rapidly rotating black hole ($a_{\rm k}=0.99$).

(4) We identify the effective region of the parameter space in $\lambda - {\cal E}$ plane that permits the shock transition in relativistic, semi-relativistic and non-relativistic flows. We observe that shock induced global accretion solutions are not stray solutions, instead they continue to exist for a large range of flow parameters. Moreover, it has been shown in this paper that the shock parameter space for relativistic and semi-relativistic flows do show close matching even when the spin of the black hole is very high ($a_{\rm k} = 0.99$). But, shock parameter space obtained for non-relativistic flow does not show any overlap with the relativistic results.

Based on the above findings, we stress that 
semi-relativistic approximation could be used to study the accretion flow dynamics using the identified effective potential ($\Phi^{\rm eff}_e$).
Our claim stems from the fact that the obtained results closely match with the relativistic one as far as the transonic and shock properties are concerned. Moreover, unlike the existing gravitational potentials \citep{Chakrabarti_Khanna1992,Artemova_etal1996,Mukhopadhyay02,Chakrabarti_Mondal2006}, this potential does not suffer any limitation due to the choice of black hole spin as it works seamlessly for $a_{\rm k} \rightarrow 1$. In reality, for all practical purposes, this potential can be successfully incorporated with ease just like a Newtonian potential. In particular, it would be possible to carry out the complete study of accretion flow including non-linearities such as shock transitions even in the presence of viscous dissipation, radiative cooling and magnetic fields around extremely rotating black holes. Since the oscillations of shocks are known to exhibit the quasi-periodic oscillations (QPOs) of the emergent high energy radiations ($i.e.$, hard X-rays), and the QPO frequency is linked as $\nu_{QPO} \sim 1/t_{\rm infall}$, where $t_{\rm infall}$ refers free fall time from shock position, the origin of the high frequency QPO can be examined as shocks usually form closer to the rapidly rotating black holes. Moreover, the precise interpretations of the spectral and timing properties of the hard radiations emanating from the accretion flows, which in turn depend on shock, would be viable that subsequently would enable one to constrain the spin of the rapidly rotating black holes 
\citep[and references therein]{Aktar_etal2017}. At the end, the most important point we would like to bring to the reader's notice that our analysis enables 
one to carry out the numerical simulations of accretion flow around rapidly rotating black hole 
very easily simply by replacing (a) the existing approximate Newtonian and/or Pseudo-Newtonian potentials by more accurate potential $\Phi^{\rm eff}_{e}$ obtained through full general relativistic consideration and (b) $\rho^{-1}(dp/dr)$ by $(h\rho)^{-1}(dp/dr)$ in the radial momentum equation. In the forthcoming efforts, we would like to take up all the above tasks that will be reported elsewhere.

 



\appendix

\section{Calculation of $\frac{\lowercase{dv}}{\lowercase{dr}}\big|_{\rm \lowercase{c}}$ for relativistic flow}

The gradient of radial velocity at the critical point given by,
$$
\frac{dv}{dr}\bigg|_{\rm c} = -\frac{-B\pm\sqrt{B^2-4AC}}{2A}.
\eqno(A1)
$$
The explicit expression of $A$, $B$, and $C$ are obtained as follows:
\begin{widetext}
$$\begin{aligned}
A = &\gamma_v^2\bigg[1 + \frac{2C_s^2}{\Gamma + 1}\bigg\lbrace\frac{1}{v^2} 
- \frac{A'\Theta_{22}}{v}\bigg\rbrace\bigg],\\
B = &-\frac{2C_s^2\gamma_v^2A'}{(\Gamma + 1)v}\Theta_{11} - 
\frac{2C_s^2}{\Gamma + 1}(N_{11} + N_{12})A'\Theta_{11},\\
C = & -\left(N_{21} + N_{22} + N_{23} + N_{24} + N_{25} + N_{26}\right),\\
N_{11}= & \frac{\left(r-a_{\rm k}^2\right)}{r\Delta} + \frac{5}{2r},
N_{12}= -\frac{1}{2\mathcal{F}}\frac{d\mathcal{F}}{dr},
N_{21}= \frac{2 (r-1)}{(r-2)^2 r^2},
N_{22}= -\frac{4 a_{\rm k}\lambda \gamma _{\phi }^2}{r^3 \Delta}
-\frac{2a_{\rm k}\lambda \gamma _{\phi }^2\Delta'}{r^2 \Delta^2}
+\frac{4 a_{\rm k}\lambda \gamma _{\phi } \gamma _{\phi }'}{r^2 \Delta},\\
N_{23}= & -\frac{8 a_{\rm k}^2 \gamma _{\phi}^2}{(r-2) r^3 \Delta}
-\frac{4 a_{\rm k}^2 \gamma _{\phi }^2\Delta' }{(r-2) r^2 \Delta^2}
+\frac{8 a_{\rm k}^2 \gamma _{\phi }\gamma _{\phi }'}{(r-2) r^2 \Delta}
-\frac{4 a_{\rm k}^2 \gamma _{\phi }^2}{(r-2)^2 r^2 \Delta },\\
N_{24}=&\Omega \gamma _{\phi }^2\lambda   \frac{2 a_{\rm k}^2-(r-3) r^2\Delta'}{r^2 \Delta^2}
-\gamma _{\phi }^2\lambda \frac{2 a_{\rm k}^2-(r-3) r^2 \Omega'}{r^2 \Delta}
-2 \lambda\Omega  \gamma _{\phi }\frac{2 a_{\rm k}^2-(r-3) r^2 \gamma _{\phi }'}{r^2 \Delta}\\
&+2 \lambda \Omega \gamma _{\phi }^2 \frac{2 a_{\rm k}^2-(r-3) r^2 }{r^3 \Delta}
+\lambda\Omega  \gamma _{\phi }^2\frac{  r^2+2 (r-3) r }{r^2 \Delta},\\
N_{25}=&\frac{2 a_{\rm k} \left(2 a_{\rm k}^2-(r-3) r^2\right) \Omega  \gamma_{\phi}^2 \Delta '}
{(r-2) r^2 \Delta ^2} 
-\frac{2 a_{\rm k} \left(2 a_{\rm k}^2-(r-3) r^2\right) \gamma_{\phi }^2 
	\Omega '}{(r-2) r^2 \Delta }
-\frac{4 a_{\rm k} \left(2 a_{\rm k}^2-(r-3) r^2\right) \Omega  \gamma_{\phi } 
	\gamma _{\phi }'}{(r-2) r^2 \Delta }\\
&+\frac{2 a_{\rm k} \left(2 a_{\rm k}^2-(r-3) r^2\right) 
	\Omega  \gamma_{\phi }^2}{(r-2)^2 r^2 \Delta }
+\frac{4 a_{\rm k} \left(2 a_{\rm k}^2-(r-3) r^2\right) \Omega  \gamma _{\phi }^2}
{(r-2) r^3 \Delta }-\frac{2 a_{\rm k} \left(-r^2-2 (r-3) r\right) \Omega  
	\gamma _{\phi }^2}{(r-2) r^2 \Delta },\\
N_{26}=&\frac{2C_s^2}{\Gamma + 1}\bigg[N_{111} + N_{121} + (N_{11} + N_{12})A'\Theta_{11}\bigg],\\
N_{111}=&  -\frac{r-a_{\rm k}^2}{r^2 \Delta}-\frac{\left(r-a_{\rm k}^2\right)\Delta'}
{r \Delta^2}-\frac{5}{2 r^2}+\frac{1}{r \Delta},
N_{121}= -4 a_{\rm k}^2 r \frac{\left(a_{\rm k}^2+r^2\right) 
	\Delta '-4 r \Delta}{\left(a_{\rm k}^2+r^2\right)^4-4 a_{\rm k}^4 \Delta^2},\\
A'=&\frac{1}{\Theta} + \frac{\Gamma'}{\Gamma} -\frac{\Gamma'}{\Gamma + 1} 
- \frac{C_s^2(\Gamma + 1)}{\Gamma\Theta},
\Theta_{11}=-\frac{2\Theta}{(N + 1)}\bigg[\frac{\left(r-a^2_{\rm k}\right)}{r\Delta} 
+ \frac{5}{2r}- \frac{1}{2\mathcal{F}}\frac{d\mathcal{F}}{dr}\bigg],\\
\Theta_{22}=&-\frac{2\Theta\gamma_v^2}{(N + 1)v},
\Omega = \frac{2 a_{\rm k}+\lambda  (r-2)}{a_{\rm k}^2 (r+2)-2 a_{\rm k} \lambda +r^3},
\Gamma'=\frac{\partial \Gamma}{\partial \Theta}{~~\rm  and~}
\gamma_\phi'=\frac{\gamma_\phi^3}{2}\lambda\Omega'.
\end{aligned}$$
Here, all the quantities have their usual meaning.
\end{widetext}

\section{$\Phi^{\rm eff}_e$ for Schwarzschild black hole ($a_{\rm k}=0$)}
For Schwarzschild black hole $(a_{\rm k}=0)$, the effective potential reduces to
$$
\begin{aligned}
\Phi^{\rm eff}_e\bigg|_{a_{\rm k}=0}=&1+\frac{1}{2} \ln \left[\frac{(r-2) r^2}{r^3-\lambda ^2 (r-2)}\right],\\
=&1-\frac{1}{2}\ln \left[1-x\right],
\end{aligned}
$$
where $x=2\left(\frac{\lambda^2}{2r^2}-\frac{1}{r-2}\right)$.

For $-1\le x<1$, we get,
$$
\begin{aligned}
\Phi^{\rm eff}_e\bigg|_{a_{\rm k}=0}=&1+\frac{2^0}{1}\left(\frac{\lambda^2}{2r^2}-\frac{1}{r-2}\right)
+\frac{2^1}{2}\left(\frac{\lambda^2}{2r^2}-\frac{1}{r-2}\right)^2\\
+&\frac{2^2}{3} \left(\frac{\lambda^2}{2r^2}-\frac{1}{r-2}\right)^3
+\frac{2^3}{4} \left(\frac{\lambda^2}{2r^2}-\frac{1}{r-2}\right)^4 + ... ...,\\
\end{aligned}
$$
where the second term in the right hand side of the above equation represents the well known Paczy{\'n}sky-Wiita effective potential \citep{Paczynsky_Wiita1980}.

%
%


\begin{thebibliography}{54}
	\bibitem[\protect\citeauthoryear{Abramowicz \& Zurek}{1981}]
	{Abramowicz_Zurek1981} Abramowicz M.~A., Zurek W.~H., 1981, ApJ, 246, 314 
	
	\bibitem[\protect\citeauthoryear{Aktar, Das, \& Nandi}{2015}]
	{Aktar_etal2015} Aktar R., Das S., Nandi A., 2015, MNRAS, 453, 3414 
	
	\bibitem[\protect\citeauthoryear{Aktar et al.}{2017}]{Aktar_etal2017}
	Aktar R., Das S., Nandi A., Sreehari H., 2017, MNRAS, 471, 4806
	
	\bibitem[\protect\citeauthoryear{Artemova, Bjoernsson, \& Novikov}{1996}]
	{Artemova_etal1996} Artemova I.~V., Bjoernsson G., Novikov I.~D., 1996, ApJ, 461, 565 
	
	\bibitem[\protect\citeauthoryear{Aschenbach}{2010}]{Aschenbach2010} 
	Aschenbach B., 2010, MmSAI, 81, 319 
	
	\bibitem[\protect\citeauthoryear{Becker \& Kazanas}{2001}]{Becker_Kazanas2001} 
	Becker P.~A., Kazanas D., 2001, ApJ, 546, 429 
	
	\bibitem[\protect\citeauthoryear{Becker \& Le}{2003}]{Becker-Le03} 
	Becker P.~A., Le T., 2003, ApJ, 588, 408 
	
	\bibitem[\protect\citeauthoryear{Bhattacharyya, Minwalla, \& Wadia}{2009}]
	{Bhattacharyya_etal2009} Bhattacharyya S., Minwalla S., Wadia S.~R., 2009, JHEP, 8, 059 
	
	\bibitem[\protect\citeauthoryear{Boyer \& Lindquist}{1967}]
	{Boyer_Lindquist1967} Boyer R.~H., Lindquist R.~W., 1967, JMP, 8, 265 
	
	\bibitem[\protect\citeauthoryear{Chakrabarti}{1989}]{Chakrabarti1989} 
	Chakrabarti S.~K., 1989, ApJ, 347, 365 
	
	\bibitem[\protect\citeauthoryear{Chakrabarti}{1990}]{Chakrabarti1990} 
	Chakrabarti S.~K., 1990, Theory of Transonic Astrophysical Flows. World
	Scientific, Singapore 
	
	\bibitem[\protect\citeauthoryear{Chakrabarti \& Khanna}{1992}]
	{Chakrabarti_Khanna1992} Chakrabarti S.~K., Khanna R., 1992, MNRAS, 256, 300 
	
	\bibitem[\protect\citeauthoryear{Chakrabarti \& Titarchuk}{1995}]
	{Chakrabarti_Titarchuk1995} Chakrabarti S., Titarchuk L.~G., 1995, ApJ, 455, 623 
	
	\bibitem[\protect\citeauthoryear{Chakrabarti}{1996}]{Chakrabarti1996} 
	Chakrabarti S.~K., 1996, ApJ, 464, 664
	
	\bibitem[\protect\citeauthoryear{Chakrabarti \& Das}{2004}]{Chakrabarti-Das04}
	Chakrabarti S.~K., Das S., 2004, MNRAS, 349, 649
	
	
	\bibitem[\protect\citeauthoryear{Chakrabarti \& Mondal}{2006}]
	{Chakrabarti_Mondal2006} Chakrabarti S.~K., Mondal S., 2006, MNRAS, 369, 976 
	
	\bibitem[\protect\citeauthoryear{Chandrasekhar}{1939}]{Chandrasekhar1939} 
	Chandrasekhar S., 1939, An Introduction to the Study of Stellar Structure. Univ. Chicago
	Press, Chicago, IL 
	
	\bibitem[\protect\citeauthoryear{Chattopadhyay \& Ryu}{2009}]{Chattopadhyay-Ryu2009}
	Chattopadhyay I., Ryu D., 2009, ApJ, 694, 492 
		
	\bibitem[\protect\citeauthoryear{Cox \& Giuli}{1968}]{Cox_Giuli1968} 
	Cox J.~P., Giuli R.~T., 1968, Principles of Stellar Structure, Vol.2: Applications to Stars.
	Gordon and Breach, New York  
	
	\bibitem[\protect\citeauthoryear{Czerny \& Elvis}{1987}]{Czerny-Elvis87} 
	Czerny B., Elvis M., 1987, ApJ, 321, 305
	
	\bibitem[\protect\citeauthoryear{Das}{2007}]{Das2007} Das S., 2007, MNRAS, 376, 1659 
	
	
	\bibitem[\protect\citeauthoryear{Dihingia, Das, \& Mandal}{2018a}]
	{Dihingia_etal2018a} Dihingia I.~K., Das S., Mandal S., 2018, MNRAS, 475, 2164 
	
	\bibitem[\protect\citeauthoryear{Dihingia, Das, \& Mandal}{2018b}]
	{Dihingia_etal2018b} Dihingia I.~K., Das S., Mandal S., 2018, JApA, 39, \#6 
	
	\bibitem[\protect\citeauthoryear{Fouxon \& Oz}{2008}]{Fouxon_Oz2008} 
	Fouxon I., Oz Y., 2008, PhRvL, 101, 261602 
	
	\bibitem[\protect\citeauthoryear{Frank, King, \& Raine}{2002}]{Frank-etal2002}
	Frank J., King A., Raine D.~J., 2002, Accretion Power in Astrophysics, 3rd
	edn. Cambridge Univ. Press, Cambridge
	
	\bibitem[\protect\citeauthoryear{Fukue}{1987}]{Fukue1987}
	Fukue J., 1987, PASJ, 39, 309 
	
	\bibitem[\protect\citeauthoryear{Fukue, Tojyo, \& Hirai}{2001}]
	{Fukue-Hirai01} Fukue J., Tojyo M., Hirai Y., 2001, PASJ, 53, 555
	
	\bibitem[\protect\citeauthoryear{Fukumura \& Tsuruta}{2004}]
	{Fukumura-Tusuruta04} Fukumura K., Tsuruta S., 2004, ApJ, 611, 964
	
	\bibitem[\protect\citeauthoryear{Gou et al.}{2009}]{Gou_etal2009}
	Gou L., et al., 2009, ApJ, 701, 1076 
	
	\bibitem[\protect\citeauthoryear{Gou et al.}{2011}]{Gou_etal2011} 
	Gou L., et al., 2011, ApJ, 742, 85 
	
	\bibitem[\protect\citeauthoryear{Hawley \& Krolik}{2001}]{Hawley-Krolik01} 
	Hawley J.~F., Krolik J.~H., 2001, ApJ, 548, 348
	
	\bibitem[\protect\citeauthoryear{Ivanov \& Prodanov}{2005}]
	{Ivanov_Prodanov2005} Ivanov R.~I., Prodanov E.~M., 2005, PhLB, 611, 34 
	
	\bibitem[\protect\citeauthoryear{Kumar \& Chattopadhyay}{2014}]
	{Kumar-Chattopadhyay2014} Kumar R., Chattopadhyay I., 2014, MNRAS, 443, 3444
	
	\bibitem[\protect\citeauthoryear{Ludlam, Miller, \& Cackett}{2015}]
	{Ludlam-etal15} Ludlam R.~M., Miller J.~M., Cackett E.~M., 2015, ApJ, 806, 262
	
	\bibitem[\protect\citeauthoryear{Liang \& Thompson}{1980}]{Liang_Thompson1980} 
	Liang E.~P.~T., Thompson K.~A., 1980, ApJ, 240, 271 
	
	\bibitem[\protect\citeauthoryear{Liu et al.}{2010}]{Liu_etal2010}
	Liu J., McClintock J.~E., Narayan R., Davis S.~W., Orosz J.~A., 2010, ApJ, 719, L109 
	
	\bibitem[\protect\citeauthoryear{Lu, Gu, \& Yuan}{1999}]{Lu-etal99} 
	Lu J.-F., Gu W.-M., Yuan F., 1999, ApJ, 523, 340
	
	\bibitem[\protect\citeauthoryear{Mandal \& Chakrabarti}{2005}]{Mandal2005} 
	Mandal S., Chakrabarti S.~K., 2005, A\&A, 434, 839 
	
	\bibitem[\protect\citeauthoryear{Manmoto, Mineshige, \& Kusunose}{1997}]
	{Manmoto-etal97} Manmoto T., Mineshige S., Kusunose M., 1997, ApJ, 489, 791
	
	\bibitem[\protect\citeauthoryear{Matsumoto et al.}{1984}]{Matsumoto-elal84} 
	Matsumoto R., Kato S., Fukue J., Okazaki A.~T., 1984, PASJ, 36, 71
	
	\bibitem[\protect\citeauthoryear{Mukhopadhyay}{2002}]{Mukhopadhyay02}
	Mukhopadhyay B., 2002, ApJ, 581, 427
	
	\bibitem[\protect\citeauthoryear{Narayan, Kato, \& Honma}{1997}]{Narayan-etal97} 
	Narayan R., Kato S., Honma F., 1997, ApJ, 476, 49
	
	\bibitem[\protect\citeauthoryear{Paczy{\'n}sky \& Wiita}{1980}]
	{Paczynsky_Wiita1980} Paczy{\'n}sky B., Wiita P.~J., 1980, A\&A, 88, 23 
	
	\bibitem[\protect\citeauthoryear{Peitz \& Appl}{1997}]{Peitz_Appl1997} 
	Peitz J., Appl S., 1997, MNRAS, 286, 681 
	
	\bibitem[\protect\citeauthoryear{Rezzolla \& Zanotti}{2013}]
	{Rezzolla_Zanotti2013} Rezzolla L., Zanotti O., 2013, Oxford University Press  
	
	
	\bibitem[\protect\citeauthoryear{Riffert \& Herold}{1995}]{Riffert_Herold1995} 
	Riffert H., Herold H., 1995, ApJ, 450, 508 
	
	
	\bibitem[\protect\citeauthoryear{Shafee et al.}{2006}]{Shafee-etal06} 
	Shafee R., McClintock J.~E., Narayan R., Davis S.~W., Li L.-X., Remillard R.~A., 2006, ApJ, 636, L113
	
	\bibitem[\protect\citeauthoryear{Sarkar \& Das}{2016}]{Sarkar_Das2016} 
	Sarkar B., Das S., 2016, MNRAS, 461, 190 
	
	\bibitem[\protect\citeauthoryear{Sarkar, Das, \& Mandal}{2018}]{Sarkar-etal2018}
	Sarkar B., Das S., Mandal S., 2018, MNRAS, 473, 2415
	
	\bibitem[\protect\citeauthoryear{Schaal \& Springel}{2015}]{Schaal_etal2015} 
	Schaal K., Springel V., 2015, MNRAS, 446, 3992 
	
	\bibitem[\protect\citeauthoryear{Schaal et al.}{2016}]{Schaal_etal2016} 
	Schaal K., et al., 2016, MNRAS, 461, 4441 
	
	\bibitem[\protect\citeauthoryear{Semer{\'a}k \& Karas}{1999}]{Semerak_Karas1999}
	Semer{\'a}k O., Karas V., 1999, A\&A, 343, 325 
	
	\bibitem[\protect\citeauthoryear{Synge}{1957}]{Synge1957} 
	Synge J.~L., 1957, The Relativistic Gas, North-Holland Publishing Co., Amsterdam
	
	\bibitem[\protect\citeauthoryear{Taub}{1948}]{Taub1948} Taub A.~H., 1948, PhRv, 74, 328 
	
	\bibitem[\protect\citeauthoryear{Yang \& Kafatos}{1995}]{Yang_Kafatos1995}
	Yang R., Kafatos M., 1995, A\&A, 295, 238 
\end{thebibliography}
\end{document}